\documentclass[aps,prl,reprint,superscriptaddress]{revtex4-1}

\usepackage{graphicx}
\usepackage[version=4]{mhchem}
\usepackage{hyperref}
\usepackage{textcomp}
\usepackage{dcolumn}
\usepackage{bm}
\usepackage[T1]{fontenc}
\usepackage{mathptmx}

\begin{document}


\title{Emergent Superconductivity in Single Crystalline \ce{MgTi2O4} Films via Structural Engineering}


\author{Wei Hu}
\author{Zhongpei Feng}
\affiliation{Beijing National Laboratory for Condensed Matter Physics, Institute of Physics, Chinese Academy of Sciences, Beijing 100190, China.}
\affiliation{School of Physical Sciences, University of Chinese Academy of Sciences, Beijing 100049, China.}

\author{Ben-Chao Gong}
\affiliation{Department of Physics and Beijing Key Laboratory of Opto-electronic Functional Materials \& Micro-nano Devices, Renmin University of China, Beijing 100872, China}

\author{Ge He}
\author{Dong Li}
\author{Mingyang Qin}
\author{Yujun Shi}
\affiliation{Beijing National Laboratory for Condensed Matter Physics, Institute of Physics, Chinese Academy of Sciences, Beijing 100190, China.}
\affiliation{School of Physical Sciences, University of Chinese Academy of Sciences, Beijing 100049, China.}

\author{Qian Li}
\affiliation{Beijing National Laboratory for Condensed Matter Physics, Institute of Physics, Chinese Academy of Sciences, Beijing 100190, China.}

\author{Qinghua Zhang}
\author{Jie Yuan}\email[]{yuanjie@iphy.ac.cn}
\affiliation{Beijing National Laboratory for Condensed Matter Physics, Institute of Physics, Chinese Academy of Sciences, Beijing 100190, China.}
\affiliation{Songshan Lake Materials Laboratory, Dongguan, Guangdong 523808, China.}

\author{Beiyi Zhu}
\affiliation{Beijing National Laboratory for Condensed Matter Physics, Institute of Physics, Chinese Academy of Sciences, Beijing 100190, China.}

\author{Kai Liu}
\affiliation{Department of Physics and Beijing Key Laboratory of Opto-electronic Functional Materials \& Micro-nano Devices, Renmin University of China, Beijing 100872, China}

\author{Tao Xiang}
\affiliation{Beijing National Laboratory for Condensed Matter Physics, Institute of Physics, Chinese Academy of Sciences, Beijing 100190, China.}
\affiliation{School of Physical Sciences, University of Chinese Academy of Sciences, Beijing 100049, China.}

\author{Lin Gu}
\author{Fang Zhou}
\author{Xiaoli Dong}
\author{Zhongxian Zhao}
\author{Kui Jin}\email[]{kuijin@iphy.ac.cn}
\affiliation{Beijing National Laboratory for Condensed Matter Physics, Institute of Physics, Chinese Academy of Sciences, Beijing 100190, China.}
\affiliation{School of Physical Sciences, University of Chinese Academy of Sciences, Beijing 100049, China.}
\affiliation{Songshan Lake Materials Laboratory, Dongguan, Guangdong 523808, China.}


\date{\today}

\begin{abstract}
Spinel compounds have demonstrated rich functionalities but rarely shown superconductivity. Here, we report the emergence of superconductivity in the spinel \ce{MgTi2O4}, known to be an insulator with a complicated order. The superconducting transition is achieved by engineering a superlattice of \ce{MgTi2O4} and \ce{SrTiO3}. The onset transition temperature in the \ce{MgTi2O4} layer can be tuned from 0 to 5 K in such geometry, concurrently with a stretched $c$-axis (from 8.51 to 8.53 \AA) compared to the bulk material. Such a positive correlation without saturation suggests ample room for the further enhancement. Intriguingly, the superlattice exhibits isotropic upper critical field $H_{\mathrm{c}2}$ that breaks the Pauli limit, distinct from the highly anisotropic feature of interface superconductivity. The origin of superconductivity in the \ce{MgTi2O4} layer is understood in combination with the electron energy loss spectra and the first-principles electronic structure calculations, which point to the birth of superconductivity in the \ce{MgTi2O4} layer by preventing the Ti-Ti dimerization. Our discovery not only provides a platform to explore the interplay between the superconductivity and other exotic states, but also opens a new window to realize superconductivity in the spinel compounds as well as other titanium oxides.
\end{abstract}


\maketitle

It is fascinating to explore new superconductor in a family of materials with rich functionalities, which not only assists in unraveling the interplay between superconductivity and other exotic states, but also enriches diverse applications. The spinel family has demonstrated rich functionalities, such as multiferroic effects \cite{hemberger2005relaxor}, anomalous magnetotransport \cite{doi:10.1002/adma.201805260}, large magnetostriction \cite{PhysRevLett.98.147203}, oxygen evolution catalysis \cite{doi:10.1021/jacs.8b13701}, and high-profile electrode performance \cite{PhysRevMaterials.2.045403}. In addition, novel phenomena have been disclosed in some materials, e.g. heavy electron feature in \ce{LiV2O4} \cite{PhysRevLett.78.3729,PhysRevLett.85.1052}, charge ordering in \ce{AlV2O4} \cite{doi:10.1143/JPSJ.70.1456}, spin fluctuations in \ce{ZnCr2O4} \cite{lee2002emergent}, orbital glass state in $\mathrm{Co}_{1+x}\mathrm{V}_{2-x}\mathrm{O}_4$ \cite{PhysRevLett.116.037201}, and orbital ordering in \ce{CuIr2S4} \cite{radaelli2002formation} and \ce{MgTi2O4} (MTO) \cite{Radaelli_2005,doi:10.1143/JPSJ.71.1848,PhysRevLett.92.056402,PhysRevLett.94.156402}. Nevertheless, so far only \ce{LiTi2O4} (LTO) has been reported as a superconductor among the spinel oxides, which was discovered half a century ago \cite{JOHNSTON1973777}. Subsequent researches on LTO gradually unveiled the novel superconductivity of spinel oxides, like the orbital-related state above the superconducting transition and the anomalies of upper critical field \cite{doi:10.1111/j.1151-2916.1999.tb02245.x,jin2015anomalous,jia2018effects,wei2018anomalies}. However, further comprehensive study requires its counterparts. Discovering more spinel oxide superconductors has always been challenging, but may open a new mine comparable to the families of copper-oxide and Fe-based superconductors \cite{RevModPhys.82.2421,doi:10.1021/ja800073m}.

\begin{figure}[h]
\includegraphics[width=0.4\textwidth]{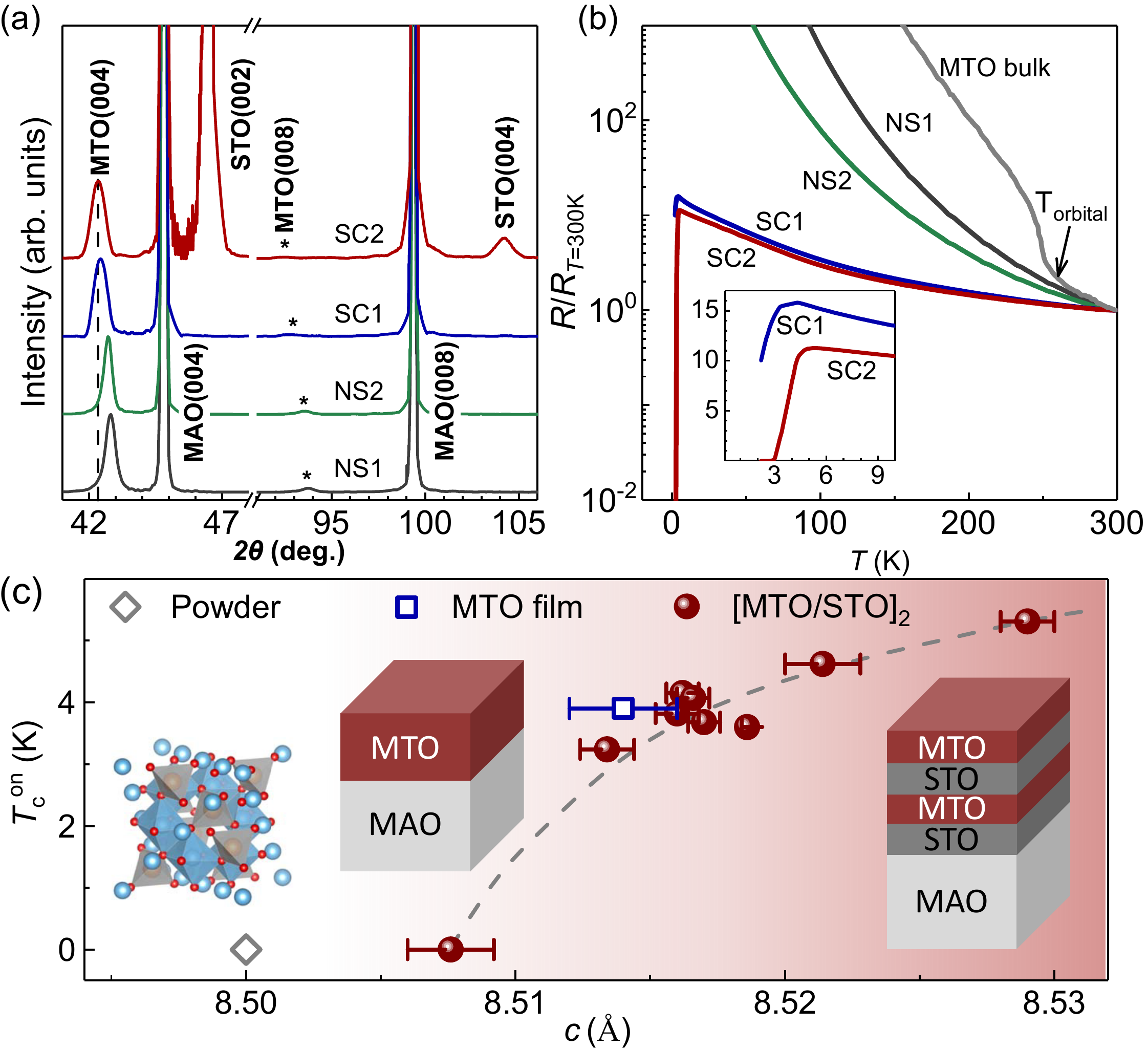}
\caption{\label{Fig1} (a) XRD spectra of $\theta$-$2\theta$ scanning. The data are shifted vertically for clarity. Samples named as NS1, NS2 and SC1 are selected MTO film samples directly grown on MAO substrate with different deposition temperature from 750{\textcelsius} to 820{\textcelsius}. The sample SC2 is a [MTO/STO]$_2$ superlattice with two periods of STO and MTO layers on MAO. (b) Temperature dependence of normalized resistance for NS1, NS2, SC1, SC2 and the MTO polycrystal \cite{ZHU2016248}. $T_{\mathrm{orbital}}$ represents the temperature at which the long-range orbital ordering steps into the MTO bulk. Inset: Zoomed-in low temperature resistance of SC1 and SC2 samples. (c) The onset superconducting transition temperature versus $c$-axis lattice of MTO. The data points, i.e. open diamond, open squares and filled circles, are extracted from the polycrystal, the MTO films and the superlattices, respectively. Insets from left to right: MTO unit cell, schematic structures of thin film and superlattice. The film thickness is about 80 nm in single layer configuration. In the superlattice, the thickness of each MTO layer and STO layer is set to 40 nm.}
\end{figure}

Since there are many similarities between MTO and LTO, e.g. approximate ionic radius between \ce{Mg^{2+}} and \ce{Li^+}, one would expect that the MTO might also show superconductivity. Considerable efforts have been made in past several decades, for instance, adjusting the Mg/Ti ratio or substituting Mg by La \cite{HOHL1996216,ISOBE200485,PhysRevB.72.045118,ZHU2016248}. However, it has not been successful in turning MTO into a superconductor. Indeed, the bulk MTO undergoes a simultaneous metal-to-insulator and cubic-to-tetragonal transition on cooling at 260 K \cite{Radaelli_2005,doi:10.1143/JPSJ.71.1848,PhysRevLett.92.056402,PhysRevLett.94.156402}. Meanwhile, orbital ordering, resulting from the helical dimerization pattern of alternating short and long Ti-Ti bonds, has been formed, responsible for the band-insulator nature of the low-temperature tetragonal phase \cite{PhysRevLett.92.056402}. Theoretical works have also mentioned that the existence of valence-bond crystal \cite{PhysRevLett.93.077208} and orbitally induced Peierls state \cite{PhysRevLett.94.156402} in MTO. Thus, it seems unrealistic for the appearance of superconductivity in such a ``robust'' insulator. Alternatively, single crystal MTO is highly desirable for a better control of the chemical composition, but challenged by the thermodynamic instability of the crystal lattice. Recalling that single crystalline LTO samples could be stabilized in the form of thin film \cite{jin2015anomalous}, we envisage that single crystalline MTO could be obtained as deposited on a lattice-match single crystal. Moreover, the compressive strain in MTO film provided by the substrate gives rise to the elongation of $c$-axis, which may suppress the dimerization as well as the orbital ordering. It is worth noting that the suppression of ordered states, e.g. antiferromagnetic order in cuprates \cite{RevModPhys.82.2421}, nematic order in Fe-based superconductors \cite{fernandes2014drives}, and charge-density-wave in $\mathrm{Cu}_x\mathrm{TiSe}_2$ \cite{morosan2006superconductivity}, is an effective approach for discovering superconductors. Therefore, the anticipation of the emergent superconductivity in MTO film is reasonable. As shown in this work, such idea is finally accomplished and the resistance can be greatly reduced in our single crystalline MTO films by adjusting the growth parameters. Remarkably, there is an indication of superconductivity starting at $\sim$ 3 K in the less resistive sample. To attain a full superconducting transition, we further engineer the MTO via the geometry of superlattice architecture, composed of MTO and \ce{SrTiO3} (STO) layers (Inset of Fig. \ref{Fig1}(c)). Eventually, the transition to zero resistance is realized, with the best onset temperature ($T_{\mathrm{c}}^{\mathrm{on}}$) up to 5 K.

Our MTO thin films and MTO/STO superlattices were grown on $[00l]$-oriented \ce{MgAl2O4} (MAO) substrates by the pulsed laser deposition (PLD) method. The STO and MTO layers were deposited by ablating a home-made \ce{SrTiO3} target and a commercial \ce{MgTi2O5} target (for that the \ce{MgTi2O4} target is unstable) using a KrF excimer laser (wavelength of 248 nm). All the samples were grown in high vacuum better than $1\times 10^{-6}$ Torr, with the pulse energy of $\sim$ 250 mJ and repetition rate of 4 Hz. The deposition temperature was set in a range from 750{\textcelsius} to 820{\textcelsius}. The crystal structures for all the samples were examined on a commercial X-ray diffractometer. X-ray diffraction (XRD) results reveal that the MTO is well oriented along the $[00l]$ direction either on the MAO substrate or adjacent to the STO layer in the superlattice.  The thickness of each layer was examined by scanning transmission electron microscopy (STEM) from the cross-section images. Transport and magnetization measurements were performed in physical property measurement system (magnetic field up to 9 Tesla) and magnetic property measurement system (remnant field less than 4 mOe), respectively.

For the MTO film on MAO, the $c$-axis lattice constant could be tuned from $\sim$ 8.44 {\AA} to $\sim$ 8.52 \AA, as manifested by the obvious shift of (004) Bragg peak to lower angle from NS1 to SC1 samples in Fig. \ref{Fig1}(a). Such evolvement is mainly caused by varying the deposition temperature. Following a stretched $c$-axis parameter, the MTO films become more conductive. Finally, an abrupt drop of the resistance is observed around 3 K (sample SC1) with decreasing the temperature as shown in Fig. \ref{Fig1}(b). This is an indication of superconducting transition, albeit not complete in the resistance. In such case, the superconducting regions in the film are not connected and a clean Meissner effect cannot be observed in the magnetization measurement. Nevertheless, it suggests that further increasing the $c$-axis may enhance the $T_{\mathrm{c}}^{\mathrm{on}}$ and realize a complete superconducting transition. Such idea is accomplished by {\it in situ} growing STO and MTO layers alternately to form a [MTO/STO]$_2$ superlattice, in which the $c$-axis lattice parameter of MTO in superlattices could be tuned in a wider range up to 8.53 \AA. For the sample with the largest $c$-axis value (SC2 in Fig. \ref{Fig1}(b)), zero resistance can be achieved at 3 K and the onset transition temperature is up to 5 K. Remarkably, the collected $c$-axis lattice values demonstrate an unambiguous link to the emergent superconductivity in our films and superlattices (Fig. \ref{Fig1}(c)). It turns out that a threshold value close to $\sim$ 8.51 {\AA} can trigger the superconducting transition in MTO, which used to be insulating by the orbital ordering. Such a positive correlation between the $c$-axis parameter and the $T_{\mathrm{c}}^{\mathrm{on}}$ has also been found in the high-$T_{\mathrm{c}}$ superconductors, such as in (Li,Fe)OHFeSe \cite{doi:10.1021/ja511292f} and in $\mathrm{La}_{2-x}\mathrm{Sr}_x\mathrm{CuO}_4$ heterostructures \cite{doi:10.1002/adma.200803850}. We also notice that the orbital ordering in bulk MTO is seriously suppressed in our single crystalline films and superlattices, evident from the smeared kink in the resistance curve (Fig. \ref{Fig1}(b)).

\begin{figure}[h]
\includegraphics[width=0.4\textwidth]{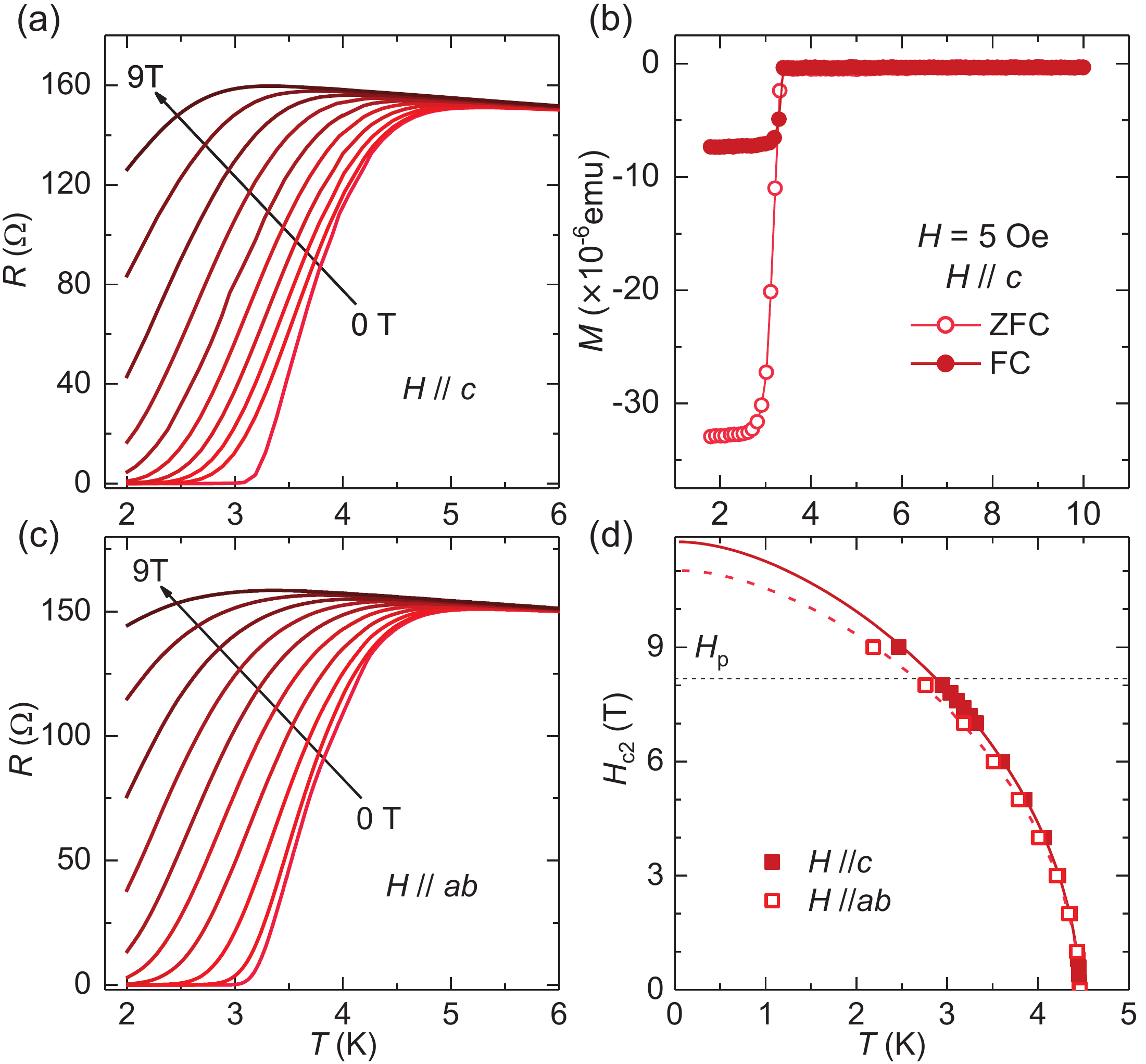}
\caption{\label{Fig2}(a) Temperature dependence of electrical resistance of [MTO/STO]$_2$ superlattice under various magnetic fields perpendicular to the $ab$-plane. (b) Temperature dependence of magnetization of [MTO/STO]$_2$ superlattice at 5 Oe with and without field cooling. (c) Temperature dependence of electrical resistance of [MTO/STO]$_2$ superlattice under various magnetic field parallel to the $ab$-plane. (d) Temperature-dependent upper critical field $H_{\mathrm{c}2}$ of the [MTO/STO]$_2$ superlattice with $H\parallel c$ (solid squares) and $H\parallel ab$ (open squares). The value of $H_{\mathrm{c}2}$ is evaluated at 90\% of the resistance transition relative to the normal state resistance. Solid and dashed lines are fits by the WHH theory. The Pauli limit $H_{\mathrm{p}}$ is marked by the dashed horizontal line.}
\end{figure}

To confirm that the zero resistance indeed comes from a superconducting transition, we also carried out magnetotransport and magnetization measurements. Fig. \ref{Fig2}(a) clearly shows that the magnetic field, applied along the $c$-axis direction ($H\parallel c$) from 0 to 9 Tesla, gradually kills the superconductivity. Meanwhile, temperature dependence of magnetic susceptibility in both zero-field-cooling and field-cooling modes discloses that the Meissner state appears at 3 K, consistent with the zero-resistance transition temperature (Fig. \ref{Fig2}(b)). The emergence of superconductivity in superlattice or multilayer structures usually shows anisotropic $H_{\mathrm{c}2}$, which is used to support a lower dimensional superconductivity confined at the interfaces, e.g. in \ce{LaAlO3}/STO \cite{Reyren1196,PhysRevLett.104.126802},  in surface superconducting \ce{MoS2} by ionic liquid gating \cite{Lu1353}. So, resistance in magnetic fields along the $ab$-plane ($H\parallel ab$, i.e. perpendicular to the $c$-axis) is also measured as shown in Fig. \ref{Fig2}(c). Surprisingly, the magnetoresistance shows similar behavior to the case of $H\parallel c$, pointing to a roughly isotropic $H_{\mathrm{c}2}$. As seen in Fig. \ref{Fig2}(d), the temperature dependence of $H_{\mathrm{c}2}$ can be well fitted by the Werthamer-Helfand-Hohenberg (WHH) theory if spin paramagnetism and spin-orbit interaction are taken into consideration \cite{PhysRev.147.295}. The $H_{\mathrm{c}2}$ at zero temperature limit is between 11 and 12 Tesla, well above the Pauli limit ($H_{\mathrm{p}}=1.84T_{\mathrm{c}}$, given by the weak-coupling BCS paramagnetic formula, here $T_{\mathrm{c}}$ is defined as the temperatures where the resistance drops to the 90\% of the normal state resistance). Breaking the Pauli limit has been constantly observed in (quasi) two-dimensional systems, which can provide key information on the nature of pairing breaking and ordering symmetry, but the $H_{\mathrm{c}2}$ is highly anisotropic \cite{PhysRevLett.104.126802,Lu1353}. However, an isotropic $H_{\mathrm{c}2}$ in a centrosymmetric system that breaks the Pauli limit is rarely seen before \cite{Fischer1978}. Very recently, we observed similar feature in superconducting LTO films, in which the anomaly is attributed to spin-orbit interaction effects and the $H_{\mathrm{c}2}$ can be even doubled by manipulating the oxygen impurities \cite{wei2018anomalies}. There is a high degree of similarity between MTO and LTO, which however, suggests that such emergent superconductivity in the MTO layer is not intuitively confined to the interface, or else, anisotropic $H_{\mathrm{c}2}$ should be expected.

\begin{figure}[h]
\includegraphics[width=0.4\textwidth]{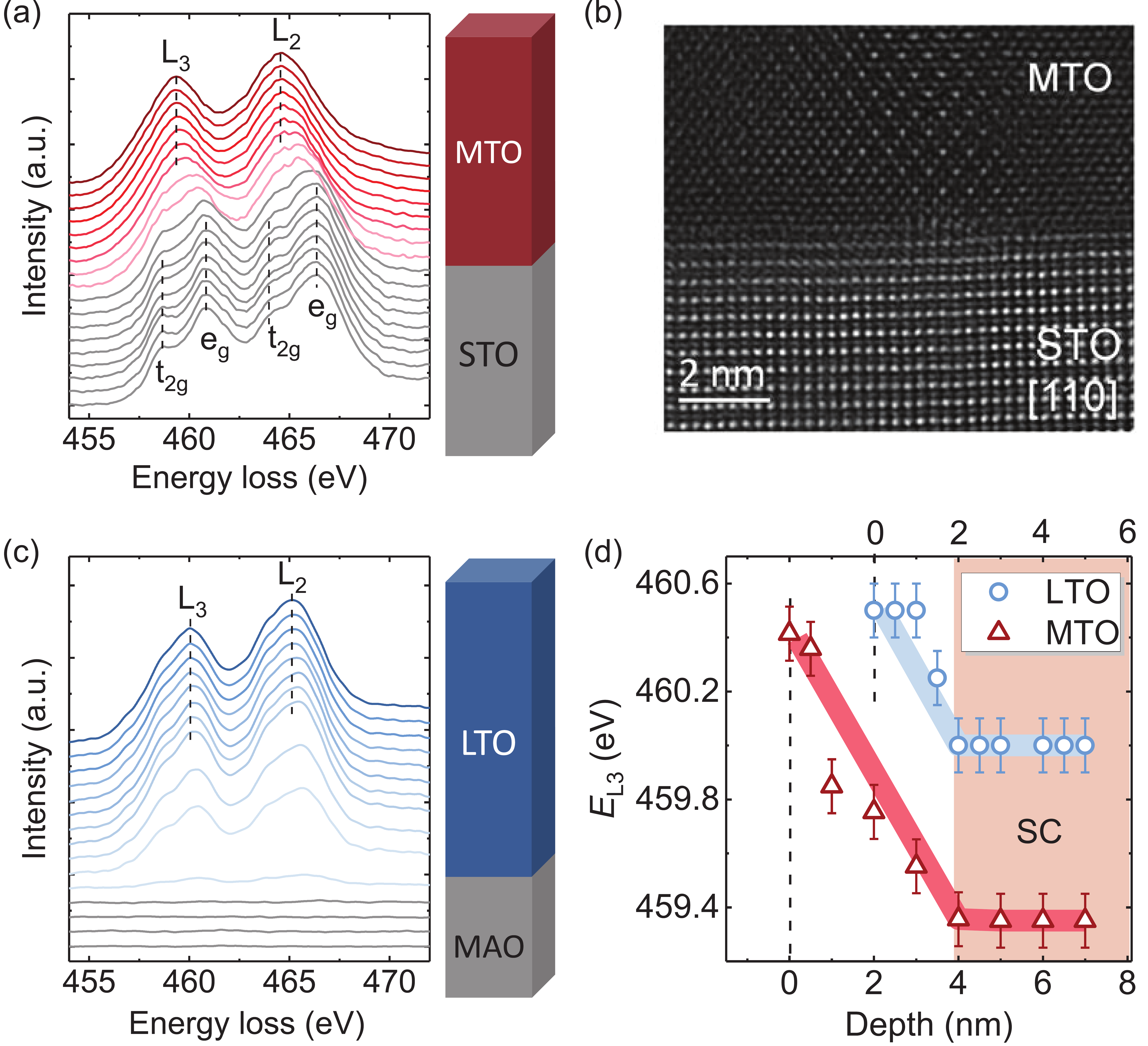}
\caption{\label{Fig3}(a) The EELS profiles for Ti $L_{2,3}$ edges of the [MTO/STO]$_2$ superlattice. The curves are shifted vertically from STO layer to MTO layer. (b) High-angle annular dark-field (HAADF) STEM image of the superlattice. (c) The EELS profiles for Ti $L_{2,3}$ edges of the LTO film on MAO substrate. (d) The depth dependence of the peak position of Ti $L_3$ edge (noted as $E_{L3}$). The data are extracted from the EELS profiles and the depth of interface is offset to 0.}
\end{figure}

Then questions arise: where does the superconductive signal come from? and what's the role of STO layer? Firstly, the STO has been reported to show superconductivity at $\sim$ 200 mK when annealed in high temperature and high vacuum \cite{PhysRevLett.12.474}. This process can remove tiny amount of oxygen and turn partial \ce{Ti^{4+}} ions into \ce{Ti^{3+}}. In our superlattice geometry, the spatial resolved electron energy loss spectroscopy (EELS) reveals that from inside the STO layer to the interface, both the Ti $L_{2,3}$ edges split into $e_{\mathrm{g}}$ and $t_{\mathrm{2g}}$ peaks and their energy positions do not vary as shown in Fig. \ref{Fig3}(a). Meanwhile, high-angle annular dark-field (HAADF) STEM image across the interface illustrates nearly prefect STO layer (Fig. \ref{Fig3}(b)). This reflects that the valence of Ti is always $+4$, no matter inside the STO layer or near the interface \cite{jia2018effects}. Besides, the reported $T_{\mathrm{c}}^{\mathrm{on}}$ of STO has not exceeded 400 mK yet \cite{ueno2008electric}. Hence, the observed superconducting signal at 5 K should not be from the STO layer. The isotropic upper critical field helps to rule out that the interface can solely be responsible for the remarkable superconductivity. Consequently, the plausibility should be locked to the MTO layer.

It is known that depositing LTO films in high vacuum can reduce the Li/Ti ratio \cite{jia2018effects}. Previous work on powders unveiled that reduced Mg/Ti ratio can result in an expansion of the crystal lattice along the $c$-axis \cite{ISOBE200485}. Therefore, the stretched $c$-axis lattice constant in our films and superlattices compared to the bulk material can be regarded as the indicator of reduced Mg/Ti ratio, yet it is hard to make a quantitative estimation. However, excessive reduction of Mg/Ti ratio will destabilize the MTO lattice. In the MTO thin film, MAO substrate helps to stabilize the crystal lattice as that in LTO experiments, in which the film is more stable than the bulk \cite{jin2015anomalous,jia2018effects}. In the [MTO/STO]$_2$ superlattice, the STO layer seems to play a role in stabilizing the MTO lattice with further reduced Mg/Ti ratio. Comparing the EELS spectra of MTO to that of STO layers, we find that it mimics the comparison between the superconducting \ce{LiTi2O4} and the insulating \ce{Li4Ti5O12} \cite{jia2018effects}. With varying distance from interface, the peak position of Ti $L_3$ edge (noted as $E_{L3}$), extracted from the high-resolution EELS spectra, firstly shifts towards lower energy and then saturates at 459.3 eV (Fig. \ref{Fig3}(d)). Intriguingly, as shown in Fig. \ref{Fig3}(c), the similar evolution is also detected in LTO film on MAO substrate where redshift of Ti $L_3$ edge is associated with $c$-axis expansion, caused by the variation of oxygen concentration. In some sense the superlattice structure assists in engineering an electronic structure of the MTO analog to the superconducting LTO. Therefore, it is intuitive to link the decrease of $E_{L3}$ in MTO with the stretch of $c$-axis. Combined with the relation of $c$-axis and $T_{\mathrm{c}}^{\mathrm{on}}$ illustrated in Fig. \ref{Fig1}(c), we speculate that the superconducting area locates in the broad region marked by shadow rather than the limited region adjacent to interface, which results in the isotropic $H_{\mathrm{c}2}$ and the absence of orbital ordering in transport results. However, we should point out that with varying the $c$-axis lattice constant, the $T_{\mathrm{c}}^{\mathrm{on}}$ of MTO is tunable compared to a robust $T_{\mathrm{c}}^{\mathrm{on}}$ in LTO, which provides plenty of room for further enhancing the superconductivity in other superlattice architectures and exploring the interplay between the superconductivity and other degrees of freedom such as the orbital ordering and spin-orbit coupling.

\begin{figure}[h]
\includegraphics[width=0.38\textwidth]{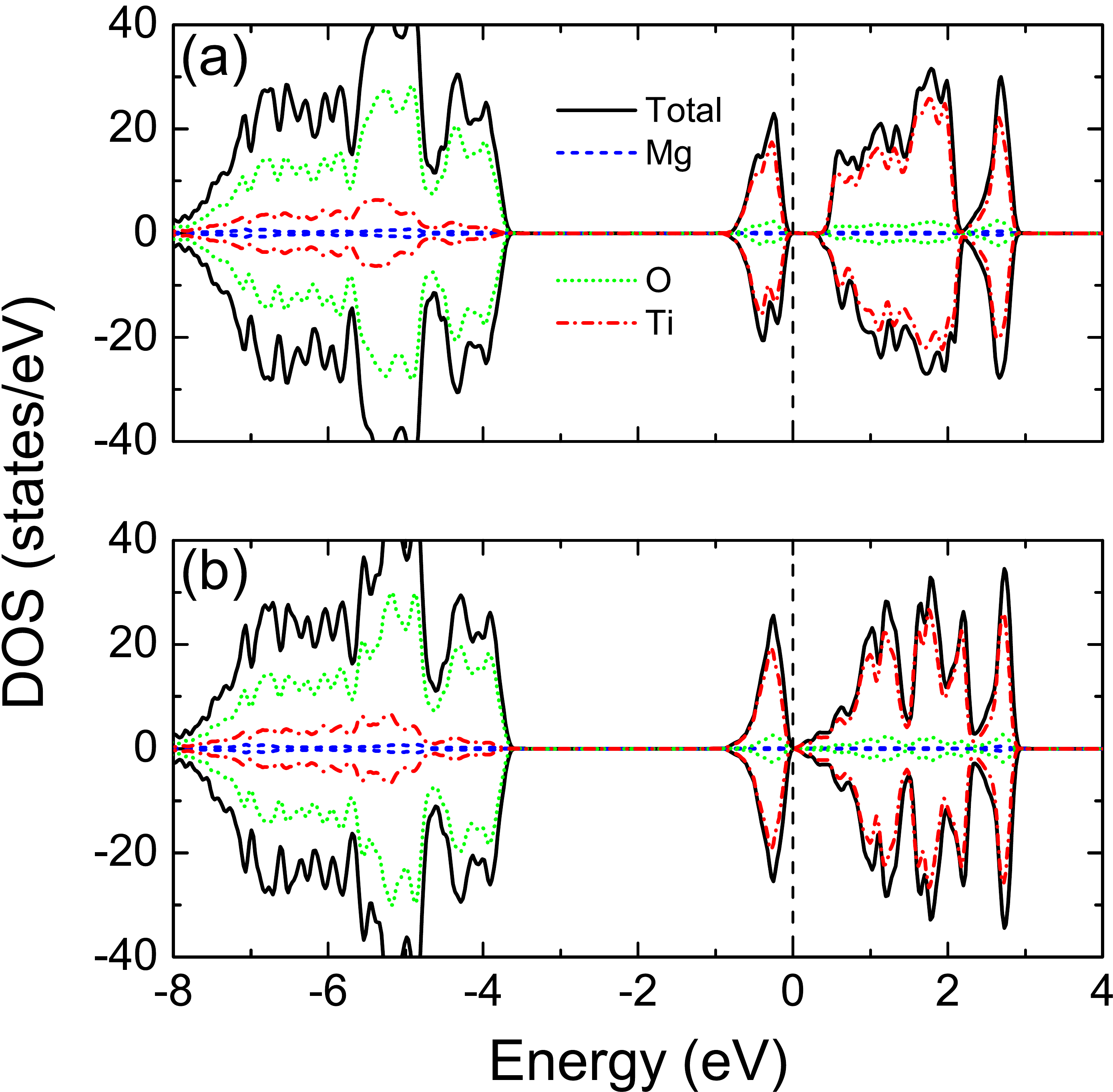}
\caption{\label{Fig4}Total and local density of states (DOS) for the antiferromagnetic state of \ce{MgTi2O4} with Hubbard $U$ of 2 eV: (a) $c$-short phase and (b) $c$-long phase. The Fermi level sets to zero.}
\end{figure}

To understand the change of electronic properties with the length of $c$-axis in MTO (Fig. \ref{Fig1}), we performed the spin-polarized first-principles electronic structure calculations by using the projector augmented wave (PAW) method \cite{PhysRevB.50.17953,*PhysRevB.59.1758} as implemented in the VASP package \cite{PhysRevB.47.558,*Kresse_1994,*PhysRevB.54.11169,*KRESSE199615}. The generalized gradient approximation (GGA) of Perdew-Burke-Ernzerhof \cite{PhysRevLett.77.3865} was employed for the exchange-correlation functional. The kinetic energy cutoff of the plane-wave basis was set to be 520 eV. A $6\times 6\times6$ $k$-point mesh was adopted for the Brillouin zone sampling of the conventional cell. The GGA+U formalism of Dudarev {\it et al.} \cite{PhysRevB.57.1505} was used  to include the correlation effect among Ti 3$d$ electrons. The effective Hubbard $U$ of 2 eV was chosen by comparing with the experimental energy gap \cite{PhysRevB.74.245102}. In structural optimization, initial atomic positions were set to experimental values \cite{PhysRevLett.92.056402} and then all atoms were allowed to relax until the forces were smaller than 0.01 eV/{\AA}. In reference to the room-temperature cubic phase with the experimental lattice constants of $a=b=c=8.509$ {\AA} \cite{PhysRevLett.92.056402}, we considered two tetragonal cells. One owns a compressed $c$ of 8.479 {\AA} (denoted by $c$-short) as the dimerized phase of bulk MTO found in previous experiment \cite{PhysRevLett.92.056402}, the other is with an elongated $c$ of 8.539 {\AA} (denoted by $c$-long) as the present film form (Fig. \ref{Fig1}(c)). The volumes of both tetragonal cells are the same as the cubic phase \cite{PhysRevLett.92.056402}, while their in-plane lattice constants are adapted accordingly. The calculated density of states for the $c$-short and $c$-long tetragonal cells are shown in Figs. \ref{Fig4}(a) and \ref{Fig4}(b), respectively. Obviously, the former exhibits the semiconducting behavior with a band gap about 0.42 eV, closed to the experimental value \cite{PhysRevB.74.245102}. It should be noted that one would obtain a metallic state for the $c$-short phase unless the Hubbard interaction was included in the calculation. This indicates that the $c$-short phase is a Mott insulator, in accordance with the result of Matteo {\it et al.} \cite{PhysRevLett.93.077208}. On the other hand, with the elongated $c$-axis (Fig. \ref{Fig4}(b)), the band width of the states right above the Fermi level ($E_{\mathrm{F}}$) broadens, giving rise to a diminished band gap. The above calculation results suggest that the $c$-axis elongation in MTO film could close the band gap and increase the density of states at $E_{\mathrm{F}}$, which are beneficial for the appearance of superconductivity.

Overall, stable single crystalline \ce{MgTi2O4} films have been successfully grown on the \ce{MgAl2O4} substrate, with the best conductive sample showing an indication of superconducting transition at $\sim$ 3K. As engineered in a geometry of [MTO/STO]$_2$ superlattice, the superconducting transition with the $T_{\mathrm{c}}^{\mathrm{on}}$ of 5 K is realized in the MTO. This emergent superconductivity can be ascribed to the elongation of $c$-axis, which is also consistent with the results of the first-principles electronic structure calculations. Except for opening a new window to accomplish more spinel oxide superconductors, such discovery also generates several attractive issues: firstly, the superlattice shows an intriguing isotropic upper critical field up to 11 Tesla that breaks the Pauli limit, distinct from the highly anisotropic feature of interface superconductivity as reported in the $\mathrm{La}_{2-x}\mathrm{Sr}_x\mathrm{CuO}_4$ heterostructure \cite{gozar2008high}, the \ce{LaAlO3}/STO \cite{Reyren1196,PhysRevLett.104.126802}, and the ultrathin FeSe film on STO \cite{Wang_2012}. So it is worthy of considering the role of STO layer in promoting the superconductivity from a new perspective. Secondly, there is a positive correlation between the $T_{\mathrm{c}}^{\mathrm{on}}$ and the $c$-axis lattice constant, so where is the upper limit and can it break the $T_{\mathrm{c}}^{\mathrm{on}}$ record of LTO in the spinel oxides? Thirdly, besides the emergent superconductivity in MTO in spinel structure, there are also reports of enhanced superconductivity in the perovskite \ce{SrTiO3} with \ce{TiO6} octahedron \cite{ueno2008electric} and in cubic titanium monoxide with wrinkled TiO plane \cite{zhang2017enhanced}, so can we find a common thread to achieve high-$T_{\mathrm{c}}$ superconductivity in titanium oxide family? It is highly anticipated to open a new mine comparable to the families of copper-oxide and Fe-based superconductors, and the emergence of superconductivity by engineering the spinel family provides the inspiration.

We would like to thank J. P. Hu, and X. H. Chen for helpful discussions. This work was supported by the Strategic Priority Research Program of Chinese Academy of Sciences (XDB25000000), the National Key Basic Research Program of China (2015CB921000, 2017YFA0303003, 2017YFA0302902, 2017YFA0302903 and 2018YFB0704102), the National Natural Science Foundation of China (11674374, 11774424, 11804378 and 51672307), the Key Research Program of Frontier Sciences, CAS (QYZDB-SSW-SLH008, QYZDY-SSW-SLH001 and QYZDB-SSW-JSC035), CAS Interdisciplinary Innovation Team, the Fundamental Research Funds for the Central Universities, and the Research Funds of Renmin University of China (19XNLG13). Computational resources were provided by the Physical Laboratory of High Performance Computing at Renmin University of China.

\end{document}